\documentclass[12pt]{article}
\usepackage{newtxtext,newtxmath}

\usepackage{graphicx}

\usepackage[letterpaper,margin=1in]{geometry}

\renewenvironment{abstract}
	{\quotation}
	{\endquotation}

\date{}

\makeatletter
\renewcommand{\fnum@figure}{\textbf{Figure \thefigure}}
\renewcommand{\fnum@table}{\textbf{Table \thetable}}
\newcommand{\etal}{\textit{et al.}}
\makeatother

\usepackage{scicite}
\usepackage{hyperref}
\hypersetup{hidelinks}
\usepackage{url}

\DeclareRobustCommand{\xray}{X\nobreakdash-ray}
\DeclareRobustCommand{\xrays}{X\nobreakdash-rays}

\def\scititle{
	 Ultrafast \xray{}-induced dephasing\\ of valence electrons
}
\title{\bfseries \boldmath \scititle}

\author{
        J.~Vonka$^{1}$,
        B.~Pedrini$^{1}$,
        J.~Ravnik$^{1}$,
        A.~Steppke$^{1,2,3}$,
        E.~Prat$^{4}$,
        S.~Reiche$^{4}$, \and
        D.~G.~Mazzone$^{3}$,
        P.~Anand$^{1,5}$,
        S.~Allenspach$^{3}$,
        Y.~Ekinci$^{1}$,
        U.~Staub$^{1}$,
        S.-W.~Huang$^{1}$,\and
        M.~Sander$^{1}$,
        R.~Mankowsky$^{1}$,
        H.~T.~Lemke$^{1}$,
        G.~Aeppli$^{1,5,6\ast}$,
        S.~Gerber$^{1\ast}$\and
	\small$^{1}$PSI Center for Photon Science, Paul Scherrer Institute; 5232 Villigen PSI, Switzerland.\and
    \small$^{2}$Physik-Institut, Universität Zürich; 8057 Zürich, Switzerland.\and
    \small$^{3}$PSI Center for Neutron and Muon Sciences, Paul Scherrer Institute; 5232 Villigen PSI, Switzerland.\and
	\small$^{4}$PSI Center for Accelerator Science and Engineering, Paul Scherrer Institute; 5232 Villigen PSI, Switzerland.\and    
    \small$^{5}$Laboratory for Solid State Physics and Quantum Center, ETH Zürich; 8093 Zürich, Switzerland.\and
    \small$^{6}$Institut de Physique, EPF Lausanne; 1015 Lausanne, Switzerland.\and
	\small$^\ast$Corresponding authors. Email: aepplig@ethz.ch and simon.gerber@psi.ch\and
}

\begin{document} 

\maketitle

\begin{abstract} \bfseries \boldmath
Diffract-before-destroy schemes for atomic-scale electron mapping assume that ultrashort X-ray pulses outrun radiation damage. Although atomic cores move over tens of femtoseconds, outer electrons respond much faster, compromising measuring their locations. We test this limit in elemental silicon using broadband, energy-chirped X-ray pulses to encode time with few-femtosecond resolution. Comparing a valence-sensitive structurally forbidden reflection with a simultaneously measured lattice Bragg peak reveals valence-bond dephasing on sub-20-fs timescales even at fluences where direct ionization is negligible. Our findings demonstrate that crystallography where outer electrons contribute substantially to Bragg peaks requires pulses shorter than today's $\sim$50~fs standard. They also establish a path to disentangle valence electron and lattice contributions to order and dynamics with hard \xrays{}, enabling studies across molecular biology, quantum matter and devices.

\end{abstract}

\noindent
Brilliant ultrafast \xray{} pulses produced by \xray{} free-electron lasers~(FEL) have become an established and widespread tool for determination of atomic-scale structure and dynamics ~\cite{seddon2017}. However, intense \xrays{} induce photoabsorption, leading to ionization and electronic rearrangements, inevitably altering spectroscopic and diffraction signals and, at high doses, causing permanent radiation damage. The ultrashort, femtosecond duration of \xray{} FEL pulses makes it possible to record signals from samples before the onset of damage processes. This concept of “diffract before destroy”~\cite{chapman2014} was successfully demonstrated for the structure determination of macromolecular protein crystals~\cite{chapman2011}. Since then, it has become a central approach in serial femtosecond crystallography~\cite{barends2022} and is increasingly also applied to single-particle imaging~\cite{sobolev2020}. Multiple experiments have examined the limits of fluence and pulse durations below which structural radiation damage can be outrun and, consequently, neglected. Most have been conducted in the framework of protein crystallography~\cite{lomb2011, nass2020}, focusing on very high fluences and using liquid solutions, in which collisional electron impact ionization is low as excited electrons can escape the small macromolecular crystals.

For solids, various studies have revealed ultrafast atomic displacement and structural disorder after \xray{} irradiation, mainly by \xray{}~pump -- \xray{}~probe techniques, where at high fluences both thermal and nonthermal mechanisms contribute to the observed \xray{} induced disorder~\cite{medvedev2015a, inoue2024}. The latter process can be described as an ultrafast atomic reordering that takes place due to the interatomic potential being altered after a large fraction (at least several percent) of the valence electrons is ionized~\cite{sundaram2002}. Independent of the mechanism, one finds that ionic displacement increases only after several tens of femtoseconds~\cite{inoue2022, inoue2025}. Recently, attosecond \xray{} pulses have become available at FELs~\cite{marinelli2017, malyzhenkov2020, yan2024}, approaching the natural timescale of electrons. This raises the question whether the diffract-before-destroy concept can be extended beyond the mere avoidance of structural damage, \textit{e.g.}, to probe pristine electronic states relevant for the performance of materials and devices. Here we show that the criterion for success is much more stringent, even at low fluences orders of magnitude below the permanent damage threshold, requiring pulse durations in the few to sub-femtosecond regime rather than those typically used at \xray{} FELs.

\subsection*{Forbidden reflections: A probe of valence electrons}
The electronic properties of condensed matter stem from the presence, redistribution and hybridization of valence electrons, determining the bond and band structure, and underpinning complex states such as superconductivity, ferroelectricity and charge-density waves~\cite{keimer2015, tokura2017}. These cannot be understood from atomic positions alone but are strongly correlated with the valence-electron density. Here we utilize the most important material of modern technology and most perfect crystal---elemental silicon---to directly measure the temporal evolution of the diffracted intensity from the valence and core electron subsystems simultaneously upon ultrafast \xray{} FEL irradiation.

\begin{figure}
	\centering
	\includegraphics[width=\linewidth]{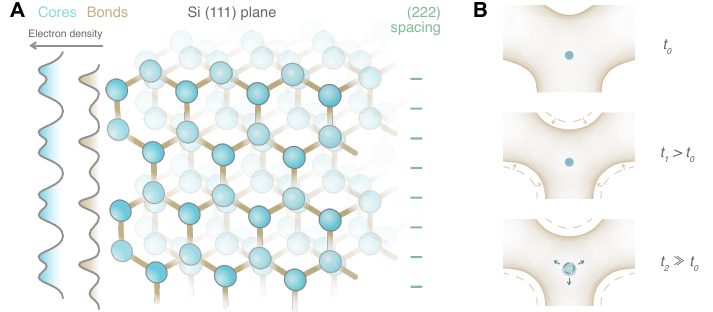} 
	\caption{\textbf{Origin and dynamics of ``forbidden'' Bragg peaks.}
		(\textbf{A})~Schematic of the silicon (111)~plane with the density profile of the core (blue) and bond (brown) electrons on the left. The (222) spacing (green lines on the right) illustrates that the symmetric core electron contribution to the structure factor completely cancels out, leaving only the asymmetric signal from the bond electrons. (\textbf{B})~Illustration of the ultrafast electronic and structural processes probed in the experiment (exaggerated for illustration). Almost immediately after photo-ionization the spatial extent and overlap of the valence-electron density is transiently reduced (dashed brown lines), reflecting a redistribution of bond charges and manifested experimentally as a dephasing of the forbidden Bragg intensity. At later times, the lattice phonon response is present.}
        
	\label{fig:forbidden_reflection} 
\end{figure}

Silicon crystallizes in the diamond structure, consisting of two face centered cubic (fcc) sublattices displaced by a real space vector $\boldsymbol{r}_0=(\tfrac14, \tfrac14, \tfrac14)$ in units of the lattice constant. The corresponding structure factor contains an interference term $I(\boldsymbol{Q})\propto \left|1+e^{i\pi(h+k+l)/2}\right|^2$, with $(h,k,l)$ being the Miller indices of the scattering vector $\boldsymbol{Q}$. The expression vanishes for $h+k+l=4n+2$, where $n$ is an integer, as for example with the (222) Bragg reflection. Nevertheless, weak scattering signals are experimentally observed and were already reported for diamond by Bragg more than a century ago~\cite{bragg1920}. The small, nonzero intensity is attributed to: i)~a static anti-centrosymmetric net charge distribution of the electron bonds, and ii)~anisotropic thermal vibrations of the spherically symmetric core electrons (see Fig.~\ref{fig:forbidden_reflection}). At room temperature the first process prevails and only at elevated temperatures anharmonic vibrations become the dominant factor determining the scattering strengths~\cite{roberto1970, hastings1975}. As the forbidden reflections arise predominantly from covalent bonds, they represent a pristine signal originating from the valence electron distribution. They thus provide a model realization of a charge-ordered state, analogous to the electronic orders studied in more complex materials~\cite{comin2016}. Revisiting the pioneering works on forbidden reflections in near-perfect crystals with ultrafast scattering provides a system with which an ambiguity analogous to the charge-density-wave versus mass-density-wave problem can be studied: a single Bragg reflection mixes electronic charge modulation with periodic lattice distortion. Time and temperature can provide contrast by isolating the fast electronic response and tuning the anharmonic contribution.

\subsection*{Self-referenced chirped-pulse diffraction}
\begin{figure}
	\centering
	\includegraphics[width=\linewidth]{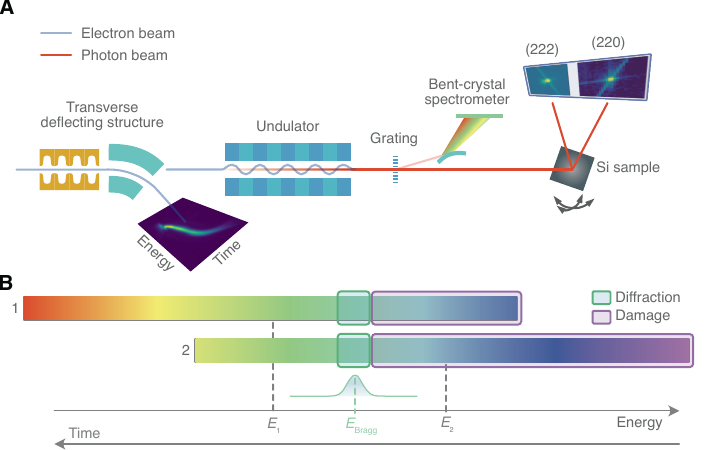}
	\caption{\textbf{Encoding femtosecond dynamics with a chirped \xray{} FEL beam.}
		(\textbf{A}) Broadband \xray{} pulses are created by the FEL through overcompression of the electron bunches. The pulse duration is then deduced from the electron beam by streaking it in a transverse deflecting structure and imaging on a screen (see Fig.~\ref{fig:chirp}). Downstream, the spectrum of each shot is measured with a bent-crystal spectrometer. In the experimental hutch, \xrays{} impinge on the silicon crystal that is aligned on a 6-circle diffractometer such that the (222)~charge and (220)~lattice Bragg conditions are fulfilled simultaneously, and the respective diffraction signals are detected at once by an area detector. (\textbf{B})~Both Bragg conditions are tuned to be fulfilled at the same energy $E_{\mathrm{Bragg}}$ (green). Photons with higher energies (purple) arrive at the sample before diffraction takes place and, consequently, do not contribute to the Bragg signals seen on the detector. Increasing the central energy of the broadband pulse, \textit{e.g.}, from $E_1$ to $E_2$, results in a larger number of photons reaching the sample before the diffraction event and, therefore, higher \xray{} induced damage and time delay.}
	\label{fig:experimental_scheme}
\end{figure}

Our approach to investigate the femtosecond dynamics of electron redistribution relies on the use of chirped hard \xray{} pulses at the SwissFEL facility~\cite{prat2020a}. A large bandwidth of \ensuremath{3\,\%} at a photon energy of $\approx\ensuremath{6\,\mathrm{keV}}$ is achieved by introducing an energy chirp to the electron beam by overcompression just prior to reaching the undulators~\cite{prat2020} (see Fig.~\ref{fig:experimental_scheme}A, as well as the Materials and Methods section). This electron energy chirp is then also imprinted onto the photon beam: high-energy \xrays{} arrive at the sample first, followed by photons with increasingly lower energies. Figure~\ref{fig:experimental_scheme}B shows the energy chirp of the electrons together with the resulting photon spectrum. Just like a monochromator, the sample acts as a bandwidth filter and selects a narrow part of the spectrum that fulfills the Bragg condition. Varying the mean energy of the electron bunch shifts the \xray{} spectrum and, since the \xrays{} with various photon energies arrive at the sample successively, changes the dose deposited on the sample before diffraction takes place.
The conversion of the energy dispersion to the time of arrival is inferred from the longitudinal properties of the electron beam prior to entering the undulators: it is streaked by a transverse-deflecting structure~\cite{craievich2013} and imaged by a scintillating screen after a dispersive element. The resulting image maps energy horizontally, via a known dispersion factor, and time vertically, via the streak calibration (see Fig.~\ref{fig:chirp}). Electron beam measurements were performed at a mean electron energy of \ensuremath{5.58\,\mathrm{GeV}}. But the preservation of the relative energy chirp~$\Delta E/E$ allows to scale it to \ensuremath{4.32\,\mathrm{GeV}} of the $\approx\ensuremath{6\,\mathrm{keV}}$ photon energy pulses used for the present dataset. This scaled electron energy matches the \xray{} spectrum from a monochromator scan at~$\approx\ensuremath{6\,\mathrm{keV}}$, confirming the correlation between the two quantities.

A bespoke scattering geometry is chosen such that the diffraction condition is fulfilled---at the same photon energy---for two different types of Bragg reflections of silicon (see Fig.~\ref{fig:experimental_scheme}A and~\ref{fig:rocking_curves}) which are recorded on an area \xray{} detector: the (220) reflection as a regular lattice Bragg peak where the core electron density contributes to scattering, and the (222) forbidden reflection associated with the valence electrons. Detecting both peaks simultaneously permits a direct, shot-to-shot comparison of the \xray{}-induced dephasing of the valence electron subsystem relative to the lattice signal. The stochastic nature of self-amplified spontaneous FEL emission~\cite{pellegrini2016} is responsible for a large energy (and consequently also arrival time) jitter among individual shots, which normally needs to be corrected with complex arrival time diagnostics and pulse normalization. Using single pulses to probe both Bragg peak signals at once eliminates these ambiguities when comparing multiple \xray{} FEL shots.

\subsection*{Valence-charge redistribution before atomic motion}
Intense hard \xray{} beams interact primarily with inner shell electrons~\cite{ziaja2005}, expelling a fraction almost immediately upon impact. The unstable core vacancy then gets quickly refilled by outer shell electrons. Energy is conserved through concurrent emission of either a photon (fluorescence) or, particularly for light elements~\cite{keski-rahkonen1974}, one or more Auger electrons. Inevitably, the expelled core and Auger electrons collide with other atom cores, leading to impact ionization and liberation of further electrons on ultrafast timescales. This creates a cascade where one \xray{} impact can result in multiple ionization events with sizable effects on scattering intensities involving bond electrons~\cite{ziaja2002}.

\begin{figure}
	\centering
	\includegraphics[width=\linewidth]{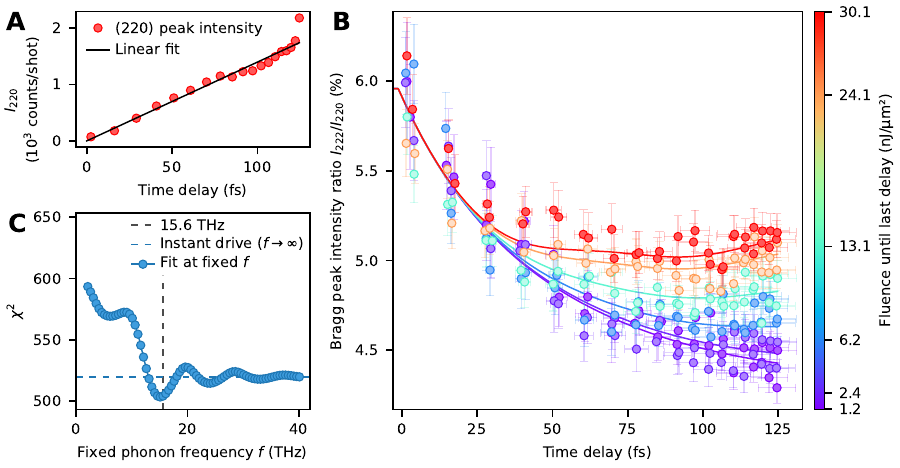}
	\caption{\textbf{Fragility of valence electrons.}
		(\textbf{A}) The total deposited X-ray dose until
the diffraction event increases continuously as $\propto t^2$ and is established from the intensity of the (220)~reflection growing as $\propto t$. (\textbf{B}) Averaged ratio of the forbidden~(222) charge and the structural~(220)~Bragg peak as a function of the time delay $t$ after the first \xray{} photon has reached the silicon crystal. The relative suppression of charge scattering directly shows that outer shell electrons are more susceptible to \xray{} induced damage. The colors denote the total accumulated \xray{} fluence $F_{\rm end}$ on the sample for the longest delay of $t=\ensuremath{125\,\mathrm{fs}}$. For each fluence, the delay was scanned in both directions without a notable difference. Solid lines are fits to the model discussed in the main text and the Materials and Methods section. Vertical error bars represent the \ensuremath{95\,\%} confidence interval of the statistics derived from 500 shots. Horizontal error bars indicate the timing uncertainty from the energy–time conversion. A global time-zero uncertainty of $\pm \ensuremath{2\,\mathrm{fs}}$ collectively shifts all data points and is not shown as a per-point error, but is taken into account as a shared time-offset fitting parameter. (\textbf{C}) Fit quality plotted as the $\chi^2$ value shows a minimum at $15.1\pm1.5\,\mathrm{THz}$, consistent with the optical phonon mode at $15.6\,\mathrm{THz}$~\cite{hase2003} (dashed black line). The limit $f\rightarrow\infty$ of an instantaneous drive is shown with a dashed blue line.}
	\label{fig:ratio}
\end{figure}

Figure~\ref{fig:ratio}A shows the temporal evolution of the (220) Bragg intensity for the highest fluence pulses ($F_{\rm end}=30.1\,\mathrm{nJ/\mu m^2}$ at the longest time delay). The approximately linear dependence is consistent with the linear increase of the incident photon flux with time, and absence of ionization of the silicon cores which would diminish the (220) signal. Matters are very different for the (222)~Bragg intensity, which we plot in Fig.~\ref{fig:ratio}B in ratio to the (220) signal,~$R=I_{222}/I_{220}$, to obtain a shot-to-shot normalization for averaging purposes. 
The data show two characteristic regimes: i)~a rapid initial suppression virtually independent of fluence in the measured range, and ii)~a divergence of traces at later times with higher fluences yielding a larger ratio~$R$. These observations suggest a phenomenological model reducing at short times only to a fluence-independent decay, and at longer times a competing, fluence-dependent process balancing the decay. We therefore describe the transient ratio as $R(t,F_{\rm end})=R_0-\Delta R_{\rm el}(t) + \Delta R_{\rm ph}(t,F_{\rm end})$, where $t$ represents the time between the arrival of the pulse front at the sample and the probing diffraction event, $F_{\rm end}$ is the incident X-ray fluence delivered by the last time delay, $R_0$ is the equilibrium Bragg-intensity ratio, $\Delta R_{\rm el}$~is the transient electronic suppression, and $\Delta R_{\rm ph}$~is the intensity restoration term attributed to a phonon response. We model the suppression as $\Delta R_{\rm el}(t)=\Delta R_0\,G(t,\tau)$, where $\Delta R_0$ is the intrinsic amplitude of the electronic suppression and $G(t,\tau)$ a normalized response function describing the intrinsic temporal evolution of the collective diffraction response with a timescale $\tau$. $G(t,\tau)$~comes from convolving the linear \xray{} pulse profile with an exponential response and accounting for the continuous arrival of photons before the diffraction event (see the Materials and Methods section for the full form, as well as the values of all fitted quantities in Tab.~\ref{tab:fitted_values}).

We attribute the restoring term $\Delta R_{\rm ph}$ to a coherent displacement of a zone-center optical phonon of silicon~\cite{hase2003}. Similar to the well-known case in bismuth~\cite{sokolowski-tinten2003}, this mode shifts the two fcc sublattices against each other along the $\left[111\right]$ direction at a frequency of $\omega/2\pi=15.6\,\mathrm{THz}$. The phonon mode changes $F_{222}$ at first order but $F_{220}$ only at second order, with the forbidden reflection being very sensitive to this mode because $F_{222}$ is a small interference residual. We define \mbox{$\Delta R_{\rm ph}(t,F_{\rm end})=A \cdot x(t)$}, where $x$ is the position of a driven harmonic oscillator \mbox{$\ddot{x}+\omega^2 x = \omega^2 F(t) / (F (t) + F_s)$} with $F(t)$ being the fluence delivered by delay $t$. The drive grows with $F(t)$ and, in our model, saturates at $F_s$. $A$~is the restoration at full saturation. A recent all-hard-X-ray transient-grating experiment on silicon observed a coherent oscillation of the (222) reflection at $16.0\pm0.7\,\mathrm{THz}$  after a ${\approx}7\,\mathrm{fs}$ long, $5\,\mathrm{keV}$ pump~\cite{ferrari2025}, supporting that hard \xrays{} can drive this mode coherently. Here the chirped pulse delivers its dose over a longer time period until the diffraction event rather than impulsively, so the mode appears on its natural timescale $1/\omega\approx10\,\mathrm{fs}$. An oscillatory response is recognizable for the strongest driven trace in Fig.~\ref{fig:ratio}B. With the frequency left free, the fit gives $15.1\pm1.5\,\mathrm{THz}$, consistent with the optical mode (see Fig.~\ref{fig:ratio}C). 

We attribute the almost immediate decrease of $R$ to an ultrafast redistribution of valence charge, resulting in a dephasing of the constructive interference from the bonding electrons that underlies the (222) diffraction signal. The globally fitted response time of $\tau=\ensuremath{18.6\pm2.1\,\mathrm{fs}}$ characterizes this collective response rather than a single electronic process. It is virtually independent of fluence and agrees with the few-femtosecond electronic response reported in optical studies of silicon~\cite{schultze2014,worle2021}. In particular, Zimin~\etal~\cite{zimin2023} measured a \ensuremath{5\,\mathrm{fs}} rise of the photoinduced absorption following a $\approx$ \ensuremath{2\,\mathrm{fs}} photoinjection and found that the Drude-Lorentz response forms within only a few femtoseconds. This finding is further supported by the selective suppression of the (222) intensity at high X-ray fluences observed by Inoue~\etal~\cite{inoue2025}. Despite assuming different photoexcitation mechanisms and spanning a large range of absorbed energy densities, all of these studies consistently identify the fast initial response with carrier redistribution to which the forbidden reflection is particularly sensitive.

Notably, the electronic sensitivity of the present measurement remains observable at absorbed energies far below the damage threshold. As shown in Fig.~\ref{fig:ratio}A, instantaneous intensity can be well-approximated as rising linearly with delay $t$, thus the deposited fluence $\propto t^2$ -- the dose at early delays is substantially lower than indicated by the colorbar in Fig.~\ref{fig:ratio}B. For instance, for $t=\ensuremath{50\,\mathrm{fs}}$, which corresponds to the time over which the suppression is already notable and the phonon response begins to be present, only \ensuremath{0.92\,\mathrm{meV/atom}} or \ensuremath{0.22\,\mathrm{nJ}/\mu\mathrm{m}^{2}} is deposited before the diffraction event for the lowest fluence used in our dataset. This is nearly three orders of magnitude below the thresholds of $\approx0.65\,\mathrm{eV/atom}$ for thermal damage and $\approx0.9\,\mathrm{eV/atom}$ for nonthermal damage~\cite{medvedev2015a}. Thus, our results demonstrate that photoinduced electron bond reconfigurations are decisive and occur already at low \xray{} fluences much below the damage threshold. Our interpretation remains phenomenological: the presented model describes the measured fluence interval and not the \mbox{$F_{\rm end}\rightarrow0$} limit in which the (222)~Bragg peak suppression must vanish. A quantitative assessment of the suppression will require an explicit excited-state calculation of $F_{222}$ with advanced non-equilibrium approaches~\cite{medvedev2018,lipp2022b}.

\subsection*{The need for attosecond X-ray pulses}

Modern semiconductor device operation also relies on the local redistribution of conduction electrons in response to internal and applied electrostatic potentials~\cite{sze2007}. Using the concept of forbidden reflections in \xray{} diffraction suppresses the core-electron background and enhances the bond-electron contrast, which paves the way to extending existing nondestructive techniques, \textit{e.g.}, for inspection of semiconductor devices~\cite{holler2019, aidukas2024}, to \textit{in operando} imaging of nanoscale structures~\cite{johannes2017}. While success has also been achieved in \xray{}-diffraction studies of exotic electronic orders predominantly involving outer-shell electrons, such as the intensely investigated charge orders of high-$T_{\rm c}$ cuprates~\cite{comin2016, ghiringhelli2012}, the results presented here raise concerns regarding the quantitative interpretation of such experiments even at moderate fluences. Moreover, the implications of our findings go beyond condensed matter physics, as biological materials, such as macromolecules, also often feature covalent bonds which break already within several femtoseconds~\cite{eliahdawod2022}. Even though not without caveats~\cite{nass2020}, structure determinations are less sensitive to these effects thanks to the delayed response of atomic displacement~\cite{suga2015}. However, damage of the valence electron subsystem becomes relevant also for such samples, \textit{e.g.} when local spin configurations or oxidation states are decisive.

The rapid suppression of \xray{} scattering from valence electrons implies that ultrashort \xray{} pulses must be used to outrun electronic damage and implement diffraction-before-destruction for the probing of pristine electronic signals. Supported by numerical simulations~\cite{lipp2026}, a solution presents itself with the attosecond pulses that have become available at \xray{} FELs~\cite{marinelli2017, huang2017, malyzhenkov2020}. Finally, techniques such as \xray{} laser-enhanced attosecond pulse generation~\cite{duris2020, guo2024} or fresh slice multi-stage amplification~\cite{wang2024}, as well as the recent demonstration of attosecond pulse generation at high repetition rates~\cite{yan2024} pave the way towards their utilization also for experiments demanding high average photon flux.


\newpage

\begin{figure}
	\centering
	\includegraphics[width=\linewidth]{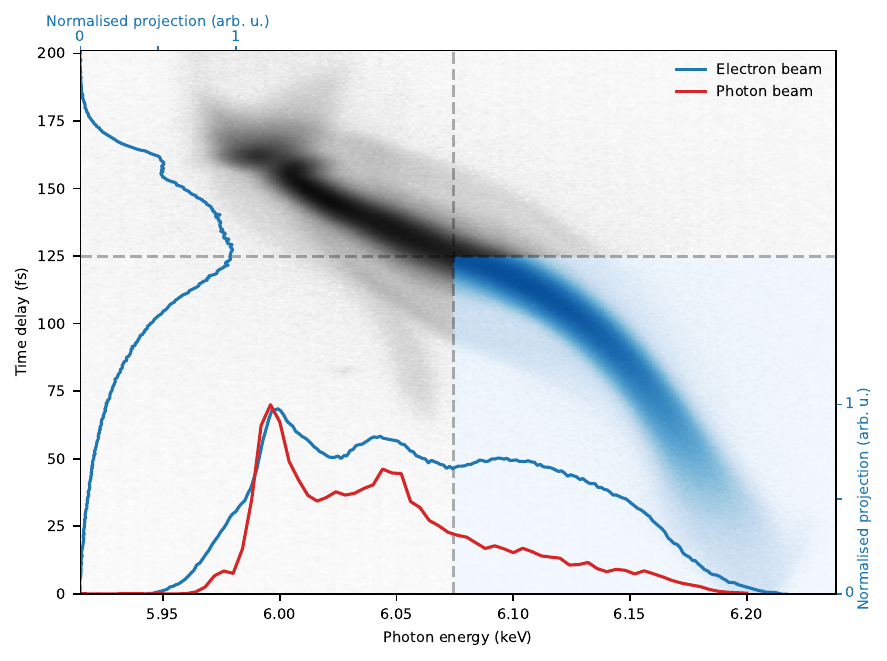}
	\caption{\textbf{Energy-to-time conversion of the chirped \xray{} FEL pulse.}
		The temporal profile of the \xray{} pulse is inferred from dispersing the electron beam, shown on the image and the projections~(blue). The horizontal units are converted from the electron to the \xray{} energy, which is confirmed with a monochromator scan of the photon beam (red; normalized to the electron beam projection). The mismatch between the electron and photon energy density at high energies originates from a reduced lasing efficiency in this region. The colored image region depicts the time-energy interval used in this work, where a roughly linear increase of the photon number is observed with decreasing energy / increasing time delay.}
	\label{fig:chirp}
\end{figure}

\clearpage 


\bibliographystyle{sciencemag}
\bibliography{Si_DbD_citations}

%
%
%
%
%
%


\section*{Acknowledgments}
We acknowledge the Paul Scherrer Institute, Villigen, Switzerland for provision of \xray{} free-electron laser beamtime at the Bernina instrument of the SwissFEL Aramis beamline and synchrotron radiation beamtime for characterization measurements at the X04SA beamline of the Swiss Light Source. We thank A.~R.~Oggenfuss, D.~Meister and M.~Lange for technical assistance during the preparation of the experiments.
\paragraph*{Funding:}
European Union’s Horizon 2020 research and innovation programme under the Marie Skłodowska-Curie Grant Agreement 701647 PSI-FELLOW-II-3i (JR). 
European Research Council under the European Union’s Horizon 2020 research and innovation programme, within Grant Agreements 810451 HERO (GA) and 681654 HyperQC (SA).
\paragraph*{Author contributions:}
Conceptualization: BP, EP, SR, GA and SG.
Experiment: All authors.
Analysis: JV, JR, AS with input from BP, EP, SR, GA and SG.
Writing – original draft: JV, GA and SG.
Writing – review \& editing: All authors.
\paragraph*{Competing interests:}
The authors declare that they have no competing interests.
\paragraph*{Data and materials availability:}
The data that support the findings of this article are openly available~\cite{Zenodo}.


\subsection*{Supplementary materials}
Materials and Methods\\
Figures S1 to S6\\
Table S1\\


\newpage


\renewcommand{\thefigure}{S\arabic{figure}}
\renewcommand{\thetable}{S\arabic{table}}
\renewcommand{\theequation}{S\arabic{equation}}
\renewcommand{\thepage}{S\arabic{page}}
\setcounter{figure}{0}
\setcounter{table}{0}
\setcounter{equation}{0}
\setcounter{page}{1} 


\begin{center}
\section*{Supplementary Materials for\\ Ultrafast X-ray-induced dephasing of valence electrons}
\author{

}
J.~Vonka, B.~Pedrini, J.~Ravnik, A.~Steppke, E.~Prat, S.~Reiche, D.~G.~Mazzone,
P.~Anand, S.~Allenspach, Y.~Ekinci, U.~Staub, S.-W.~Huang, M.~Sander, R.~Mankowsky, H.~T.~Lemke,
G.~Aeppli$^{\ast}$ and S.~Gerber$^{\ast}$\\

\small$^\ast$Corresponding authors. Email: aepplig@ethz.ch and simon.gerber@psi.ch\\

\end{center}

\subsubsection*{This PDF file includes:}
Materials and Methods\\
Figures S1 to S6\\
Table S1\\


\newpage


\subsection*{Materials and Methods}

\subsubsection*{Broadband \xray{} pulses}
The experiment was performed at the SwissFEL facility using the hard \xray{} beamline Aramis~\cite{prat2020a}. Broadband pulses with \ensuremath{190\,\mathrm{eV}} bandwidth at a photon energy of $\approx\ensuremath{6\,\mathrm{keV}}$ and a repetition rate of \ensuremath{100\,\mathrm{Hz}} were produced by applying a bunch overcompression technique~\cite{prat2020}. The electron beam entering a magnetic chicane featured a positive energy chirp, required for the compression, with the low-energy electrons at the head and high-energy ones at the tail. Normally, the direction of the chirp is preserved after the compressor and longitudinal wake fields in the following linear accelerator section remove the chirp to minimize the bandwidth. However, here the initial energy chirp was higher such that the chicane reversed the chirp, \textit{i.e.}, electrons at the head had a higher energy than the ones at the tail. In this case, the longitudinal wake fields in the linear accelerator further amplify the energy spread and increase the chirp. Different longitudinal parts of the energy-chirped electron bunch emit at distinct wavelengths in the undulator section, creating a large-bandwidth \xray{} pulse. Also in the time domain the characteristics of the electron beam are preserved: the highest energy photons arrive at the sample first, followed by the lower energy ones. A shift of the entire broadband spectrum is achieved by variation of the electron beam energy.

The temporal profile of the electron beam was measured after the last linear accelerator section, just before the undulators. There, the beam was streaked by a transverse-deflecting structure~\cite{craievich2013} and imaged on a scintillating screen as shown in Fig.~\ref{fig:chirp} of the main text. A dispersion factor, known from the accelerator layout and optics, allows to convert the horizontal position of the screen to energy, whereas the vertical one corresponds to time. The streaked electron beam was measured at \ensuremath{5.58\,\mathrm{GeV}} mean electron beam energy, while \ensuremath{4.32\,\mathrm{GeV}} was used to produce the $\approx\ensuremath{6\,\mathrm{keV}}$ photon beam for the dataset of this study. The relative energy chirp $\Delta E/E$ is preserved when the electron beam energy is varied, which allows for scaling of the temporal profile to the ultimately used electron beam parameters. The good match between the scaled electron beam spectral profile and the photon beam spectrum is shown in Fig.~\ref{fig:chirp} of the main text. To account for possible nonlinearities in the scaling, a conservative relative uncertainty of \ensuremath{5\,\%} is assigned to the energy conversion and propagated to the time axis used for the analysis.

The conversion of the electron beam energy to the \xray{} photon wavelength is governed by the FEL resonance equation~\cite{huang2007}
\begin{equation}
    \lambda = \frac{\lambda_u}{2\gamma^2} \left(1 + \frac{K^2}{2}\right),
\end{equation}
where $\lambda_u$ is the undulator period, $\gamma$ the relativistic Lorentz factor of the electron bunch and $K$ the undulator parameter. 

\begin{figure} 
	\centering
	\includegraphics[width=4.75in]{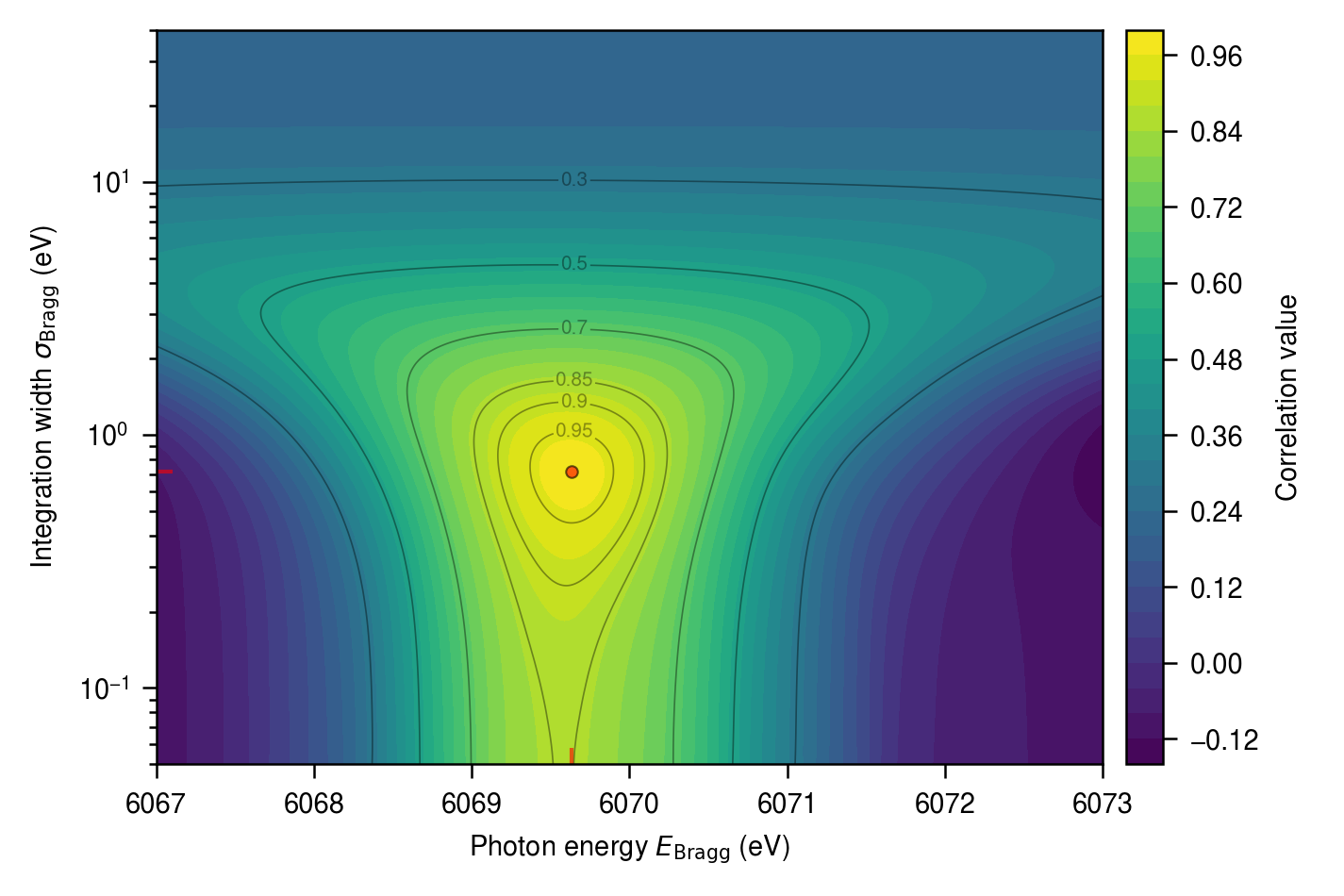} 
	\caption{\textbf{Spectral window contributing to the (220) Bragg reflection.}
    Correlation between the (220) Bragg peak intensity and the integrated signal of the upstream single-shot spectrometer as a function of the central energy  $E_{\mathrm{Bragg}}$ and the bandwidth $\sigma_{\mathrm{Bragg}}$. A high correlation value indicates the photon energy range contributing the strongest to the diffraction signal.}
	\label{fig:correlation_search}
\end{figure}

The time delay between the arrival of the first photons at the sample and the occurrence of Bragg diffraction was established from the early data points recorded by an upstream bent-crystal spectrometer~\cite{david2021} of the Aramis beamline. First, the photon energy satisfying the Bragg condition of the silicon (220) and (222) reflections $E_{\mathrm{Bragg}}$ was determined by identifying the energy window on the spectrometer having the highest intensity correlation with the Bragg peak signals. An example of this method is shown in Fig.~\ref{fig:correlation_search} for the (220) reflection, in a good agreement with measurements obtained using monochromatic beam (see Fig.~\ref{fig:rocking_curves}). Let $E_{\mathrm{max}}$ be the highest energy of the spectrum above the instrument noise floor. The energy difference $\Delta E = E_{\mathrm{max}} - E_{\mathrm{Bragg}}$ is then used to determine the nominal time delay $t$, with an uncertainty of $\sigma_{t_0}=\ensuremath{2\,\mathrm{fs}}$, when the head of the photon pulse arrives at the sample. The limited bandwidth of the spectrometer prohibits the use of this method for the entire scanned energy range. Therefore, the values of the subsequent time delays (at different central electron beam energies) are deduced from the central electron beam energy setting, taking into account the additional uncertainty from the electron-to-photon energy conversion discussed above.

In summary, the uncertainty of the time axis consists of two contributions. On the one hand, the global time-zero uncertainty $\sigma_{t_0}=2\,\mathrm{fs}$ determined from the photon spectrum, reflecting the uncertainty of the absolute arrival time of the photon pulse and leading to a rigid shift of the entire dataset. This contribution is not treated as a per-point error, but is instead included as a free parameter $t_0$ (constrained by a Gaussian uncertainty) in the fitting of Fig.~\ref{fig:ratio}B of the main text. In addition, a per-point timing uncertainty $\sigma_t$ arises from the earlier mentioned electron-to-photon energy conversion of \ensuremath{5\,\%}, which increases with time delay.

\subsubsection*{Experimental setup}

The experiment was conducted at the Bernina experimental station of SwissFEL~\cite{ingold2019}. Higher \xray{} beam harmonics were suppressed by using two bare-silicon harmonic-rejection mirrors at an angle of \ensuremath{4\,\mathrm{mrad}}. Linear horizontally polarized \xrays{} from the undulators were focused using a Kirkpatrick-Baez (KB) mirror system. Using all four mirrors at \ensuremath{4\,\mathrm{mrad}} results in virtually complete suppression of the third harmonic signal at $\approx\ensuremath{18\,\mathrm{keV}}$.
The beam spot size at the sample position was set to \ensuremath{10\times10\,\mu\mathrm{m}^{2}} as confirmed with a calibrated \xray{} microscope at the sample position using \ensuremath{10\,\mathrm{keV}} \xrays{}. At \ensuremath{6\,\mathrm{keV}}, the energy used to record the dataset of this study, the spot size was confirmed by replacing the sample by an yttrium aluminum garnet crystal and measuring the size of the fluorescent spot with a calibrated, long working distance optical microscope.

The photon count per unattenuated \xray{} pulse over the whole duration was estimated at $\approx$\ensuremath{5.6\times10^{10}\,\mathrm{photons}} based on the upstream photon beam intensity gas monitor and the expected beamline transmission of \ensuremath{30\,\%}. Though, only the early portion of the beam is used in the present analysis, as shown in Fig.~\ref{fig:chirp} of the main text. The accumulated single \xray{} pulse dose on the sample was estimated as \ensuremath{3.8\times10^{8}} and \ensuremath{9.2\times10^{9}\,\mathrm{photons/shot}} for the lowest and highest transmission, respectively, of the dataset shown in Fig.~\ref{fig:ratio} of the main text. However, the fluence until the diffraction event is substantially lower, particularly for the early delay times, as only a small fraction of the pulse contributes to the \xray{}-induced damage. The fluence estimation over the entire pulse duration was cross-checked by determining the permanent damage threshold at a different spot on the sample, and estimated to occur at $\approx1.1\times10^{10}\,\mathrm{photons/pulse}\approx0.58\,\mathrm{eV/atom}$. While this is based on a relatively coarse measurement, it is consistent with the literature value of $\approx$\ensuremath{0.65\,\mathrm{eV/atom}}~\cite{medvedev2015a}.

For measurements the central electron beam energy was varied in both directions---from low to high and high to low energies---in two subsequent scans. A minor hysteresis between two directions was observed and a correction of \ensuremath{0.29\,\mathrm{MeV}} was applied to match the two directions consistently across all measurements. The correction is within \ensuremath{12\,\%} of the step size and \ensuremath{7\times10^{-5}} of the electron energy value. Apart from this hysteresis, no systematic differences were observed among the two scan directions, indicating absence of accumulated radiation damage and removing the need for selecting a fresh probing location for consecutive measurements. Within each direction, $R$ was calculated as the ratio of the mean intensities $\langle I_{222}(t)\rangle/\langle I_{220}(t)\rangle$, rather than as a mean of shot-wise intensity ratios $\langle I_{222}(t)/ I_{220}(t)\rangle$.

\begin{figure} 
	\centering
	\includegraphics[width=6in]{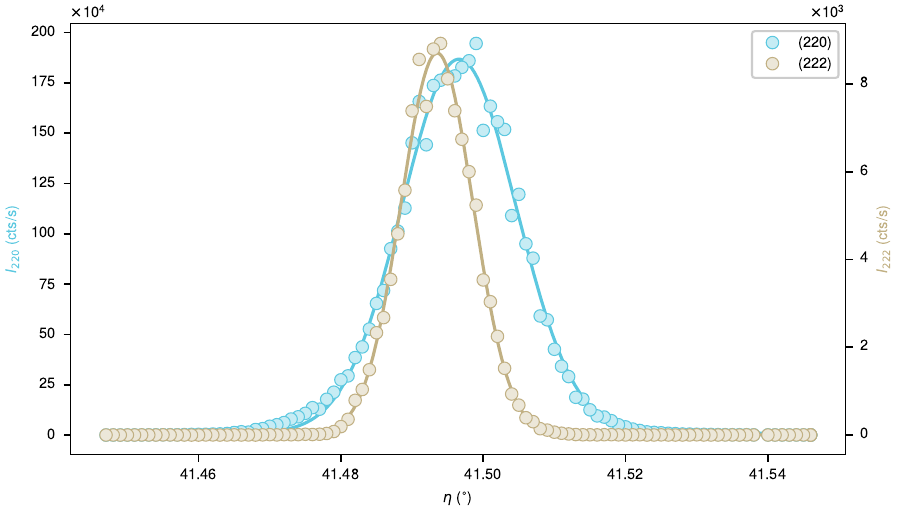} 
	\caption{\textbf{Alignment of the lattice and forbidden Bragg peaks.}
		Simultaneous scanning of the structural~(220) and the electronic~(222) Bragg peaks with monochromatized beam at a photon energy of \ensuremath{6069.1\,\mathrm{eV}}. The observed angular mismatch of \ensuremath{0.003\,{}^{\circ}} corresponds to a negligible difference of \ensuremath{0.4\,\mathrm{eV}} in photon energy.}
	\label{fig:rocking_curves}
\end{figure}

A Boron p-doped silicon single crystal (SM52 from \textit{Si-Mat}) with an estimated dopant concentration of $\sim\ensuremath{1\times10^{15}\,\mathrm{cm}^{-3}}$ and a (111) surface normal was used. The sample was oriented in the beam using a 6-circle diffractometer. All measurements were conducted at ambient conditions with nitrogen gas blowing at the sample to avoid carbon deposition at the surface. For a given photon energy, the (222) and (220) Bragg peaks were simultaneously detected on the same Jungfrau area detector~\cite{mozzanica2018}. To balance the scattered beam intensities and optimize the detector’s dynamic range, 32 and \ensuremath{160\,\mu\mathrm{m}} of aluminum foil was used to cover the area of the (222) and (220) Bragg peaks on the detector, respectively. The background intensity, calculated as the region of interest sum in the vicinity of each Bragg peak, was subtracted from the scattering intensity to rule out the influence of a possible background variation. Zero-photon noise was suppressed by using a minimum-energy threshold at each pixel at half of the single photon energy of $3\,\mathrm{keV}$. An alignment scan, using spectrally filtered \xrays{} by a Si~(111) double crystal monochromator, confirmed overlap between the two sharp Bragg peaks (see Fig.~\ref{fig:rocking_curves}), where the angular deviation was measured to be within~\ensuremath{3.2\,\mathrm{m}^{\circ}}, corresponding to a photon energy (and thus timing) mismatch of $\approx\ensuremath{0.4\,\mathrm{eV}}$---a negligible value in comparison to the \ensuremath{125\,\mathrm{eV}} energy range scanned in the dataset.

\begin{figure} 
	\centering
	\includegraphics[width=4.75in]{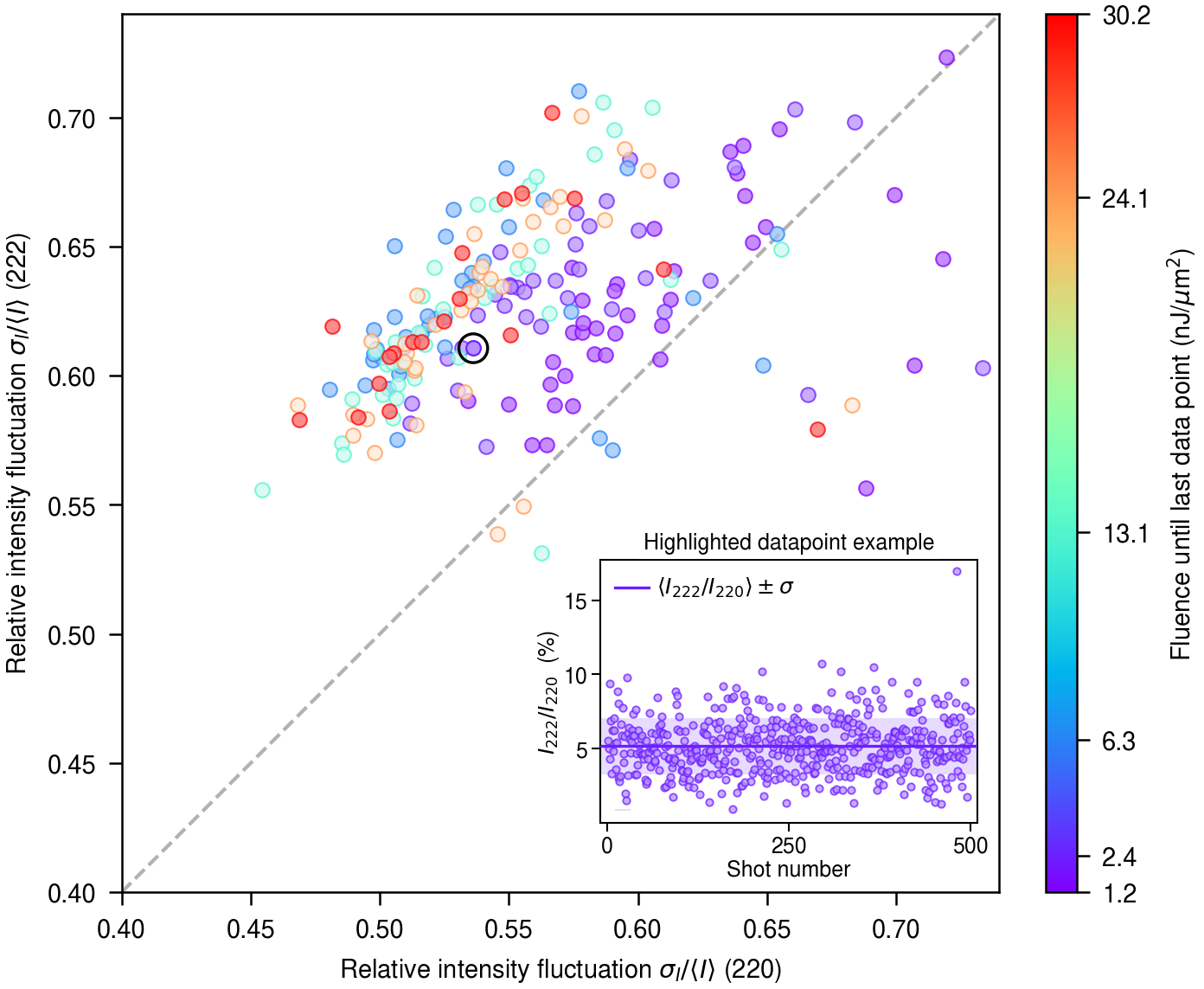} 
	\caption{\textbf{Shot-to-shot intensity fluctuations of the (220) and (222) Bragg peaks.}
The relative fluctuations, expressed as $\sigma_I / \langle I \rangle$, are shown for both reflections across the measured central photon energies (time delays) and fluences. Each datapoint corresponds to 500 shots. The gray dashed line indicates equal fluctuation amplitudes of both reflections. The absence of substantially larger fluctuations in either reflection indicates no enhanced sensitivity to geometric jitter, consistent with negligible multiple-scattering contributions. The inset shows the shot-to-shot distribution of the intensity ratio $I_{222}/I_{220}$ for a representative datapoint indicated by an open circle in the main panel.}
	\label{fig:multiple_scattering}
\end{figure}

The near-perfect crystallinity of the silicon sample results in a very narrow angular acceptance for the multiple-scattering conditions, making multiple scattering highly sensitive to geometric jitter, and thus detectable through shot-to-shot intensity analysis, which we assess by comparing the shot-to-shot intensity fluctuations of the (220) and (222) Bragg peaks. For each central photon energy (time delay) and fluence, the relative fluctuations are quantified using the coefficient of variation $\sigma_I / \langle I \rangle$, where $\sigma_I$ denotes the standard deviation and $\langle I \rangle$ the mean intensity, evaluated over 500 shots per data point (see Fig.~\ref{fig:multiple_scattering}). Fluctuations of the (222) reflections are only slightly larger, as expected for the weaker reflection, which speaks against a noticeable multiple-scattering contribution in the chosen scattering geometry. Neither were signatures of multiple scattering observed during the alignment fine-tuning of the two Bragg peaks.

As an additional consistency check,  prior to installation of the detector attenuation foils, the integrated intensity ratio of the two rocking curves in Fig.~\ref{fig:rocking_curves} was $I_{222}/I_{220}=2.88\times10^{-3}$. This value is within a factor of $\sim4$ of the structure-factor estimate $|F_{222}|^2/|F_{220}|^2 \approx 6.8\times10^{-4}$ based on the room-temperature structure factors~\cite{wahlberg2016}. The mild discrepancy is plausible given extinction and dynamical-diffraction effects for the strong (220) reflection, as well as, different penetration depths of the two reflections. In addition, there are also uncertainties in the exact attenuation correction for different, foil-covered areas of the detector. Importantly, the analysis presented here depends only on relative changes of the reflection ratio and is, therefore, insensitive to the absolute normalization. This is crucial given that the narrow widths of the two Bragg peaks did not allow for a reliable shot-by-shot normalization of the incoming beam intensity in the respective spectral range of the upstream spectrometer.

\subsubsection*{Beam trajectory}

Minimal spatial chirp is essential for a systematic measurement and ensuring that radiation damage occurs in the same region as the subsequent probing. That is, for a given central electron energy all photon energies of the spectrum should be focused on the same sample spot and this spot should remain unaffected by the variation of the electron beam energy / when changing the time delay.

\begin{figure} 
	\centering
	\includegraphics[width=4.75in]{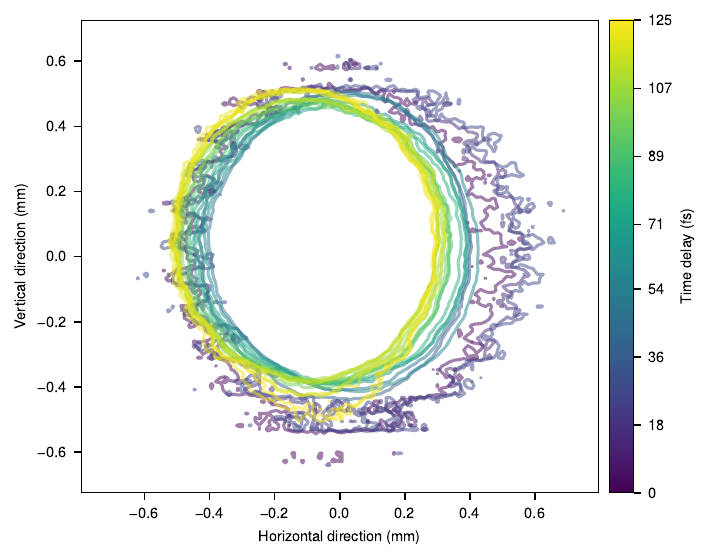} 
	\caption{\textbf{Beam position variation for a given electron beam energy.}
    Beam position and shape as measured by the beam profile monitor just before the KB focusing mirrors. Different photon energies (corresponding to different arrival times at the sample position) are selected by the upstream monochromator. The contour levels denote the mean intensity value.}
	\label{fig:beam_drift_PPRM}
\end{figure}

Beam trajectory drifts were quantified and corrected using a photon beam profile monitor positioned upstream of the sample, immediately before the focusing KB mirrors, the final optical element of the beamline. For this check the upstream monochromator was used to select specific photon energies at a constant electron beam energy. As the monochromator was scanned across the spectrum, the beam profile evolution was recorded (see Fig.~\ref{fig:beam_drift_PPRM}). A minor trajectory variation was observed in the horizontal scattering plane; however, this deviation remained well within the beam spot size, ensuring negligible impact on the measurements.

\begin{figure} 
	\centering
	\includegraphics[width=6in]{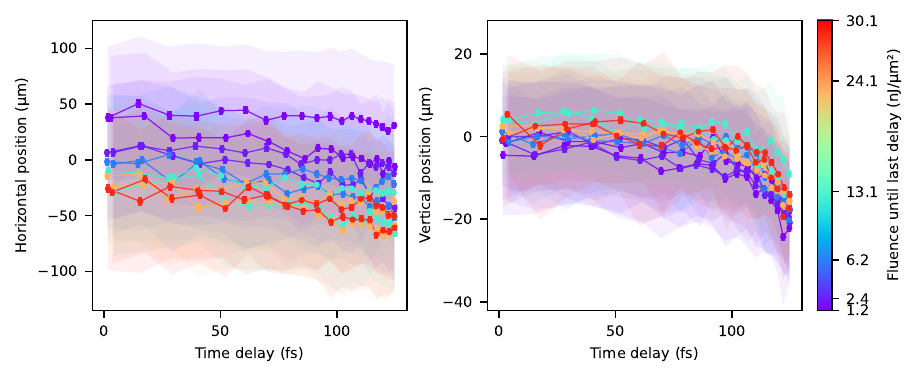}
	\caption{\textbf{Beam position variation as a function of the central electron beam energy and fluence.} Each data point is a sum of all photon energies at a given electron beam energy. The beam position was recorded by a back-scattering monitor prior to the KB focusing mirrors. Given the beam spot size of $\approx\ensuremath{0.5\,\mathrm{mm}}$ at this position, the drift is relatively small. Shaded regions indicate the one-standard-deviation, shot-to-shot beam-position jitter ($\pm 1\sigma$), while error bars show the standard error of the mean position ($\sigma/\sqrt{N}$, with 500 shots per point).}
	\label{fig:beam_drift_PBPS} 
\end{figure}

Variation of the beam trajectory at different electron beam energies was continuously monitored and corrected using a beam positioning system~(BPS). This noninvasive device relies on detecting back-scattered \xrays{} from a thin diamond wafer. The signal is recorded by four \xray{}-sensitive diodes and the beam position is determined based on the combined intensities of those~\cite{juranic2018}. Beam drifts were observed to be within \ensuremath{100\,\mu\mathrm{m}} in both horizontal and vertical directions, as illustrated in Fig.~\ref{fig:beam_drift_PBPS}. Given that the beam diameter at the BPS location (before the focusing elements) is $\approx\ensuremath{0.5\,\mathrm{mm}}$ full-width-at-half-maximum, this level of drift is relatively small in comparison to the beam size. In addition to a systematic position drift, the presence of beam jitter, originating from the stochastic self-amplified spontaneous FEL emission process is evident from the error bars. However, its magnitude remains consistent across all \xray{} energies, confirming that it does not introduce systematic errors, as each data point represents an average over 500 shots for a given fluence and scanning direction. The resulting jitter amplitude is $\approx\ensuremath{0.2\,\mathrm{mm}}$ in the vertical and $\approx\ensuremath{0.04\,\mathrm{mm}}$ in the horizontal direction. Additionally, within the angular pixel resolution of \ensuremath{52\,\mathrm{arcsec}} of the experiment, no drift of the two Bragg peaks was observed on the detector.

\subsubsection*{Phenomenological model}

As explained in the main text, we model the observed scattering ratio as
\begin{equation}
    R(t,F_{\rm end}) = R_0 - \Delta R_0 \cdot G(t,\tau) + A \cdot x(t),
\label{eq:ratio_equation_supmat}
\end{equation}
where the $\Delta R_0 \cdot G(t,\tau)$ is the scattering intensity suppression attributed to valence electrons with an intrinsic timescale $\tau$ while the term $A \cdot x(t)$ partly restores $R$ through a phonon response. Throughout this section $t$ is counted from the fitted time zero, \emph{i. e.} it stands for $t-t_0$.

The response function $G(t,\tau)$ accounts for the fact that energy is deposited continuously as the \xray{} pulse illuminates the sample, with the instantaneous intensity increasing linearly with time during the pulse, so $F(t)\propto t^2$. This approximation is justified in the investigated delay time window up to $t=\ensuremath{125\,\mathrm{fs}}$ based on the spectral profile established with the monochromator (see the red trace in Fig.~\ref{fig:chirp} of the main text), as well as the approximately linear increase in the scattering intensity of the structural (220) reflection (see in Fig.~\ref{fig:ratio}A). An excitation created at time $s$ contributes with a response proportional to $1-e^{-(t-s)/\tau}$. The measured signal at a probe delay of $t$ is, therefore, proportional to the fluence-weighted integral of these responses,
\begin{equation}
    G(t,\tau) = \frac{1}{\int_0^{t} s\,\mathrm{d}s}\int_0^{t} s\left[1-e^{-(t-s)/\tau} \right] \mathrm{d}s.
\label{eq:transfer_function_G_integral}
\end{equation}
The denominator normalizes the response to unity. Evaluating the integral yields
\begin{equation}
    G(t,\tau)=\frac{\frac{t^2}{2}-\tau^2 e^{-t/\tau}\left[(t/\tau-1)e^{t/\tau}+1\right]}{t^2/2}.
    \label{eq:transfer_function_solved}
\end{equation}

The position $x(t)$ of an undamped driven mechanical oscillator is
\begin{equation}
    \ddot x(t)+\omega^2x(t)=\omega^2x_{\rm eq}(t)\ ,\qquad x(0)=0\ ,\quad \dot x(0)=0,
\label{eq:mechanical oscillator}
\end{equation}
with the equilibrium displacement $x_{\rm eq}(t)$ driven with a saturating response $x_{\rm eq}(t) = \frac{F(t)}{F(t) + F_s}$. The equation of motion has the solution
\begin{equation}
    x(t)=\omega\int_0^t x_{\rm eq}(s)\,\sin\omega(t-s)\,ds.
    \label{eq:mechanical oscillator solution}
\end{equation}
Inserting the linearly rising intensity with a saturating drive we obtain
\begin{equation}
    x(t,F_{\rm end})=\omega\int_{0}^t \frac{F(s)}{F(s)+F_s}\,\sin\omega(t-s)\,ds.
    \label{eq:mechanical oscillator solution with drive}
\end{equation}
Unlike the response function $G(t,\tau)$, this integral does not have a solution form in elementary functions and it is thus evaluated numerically.

\begin{figure}
	\centering
	\includegraphics[width=6.0in]{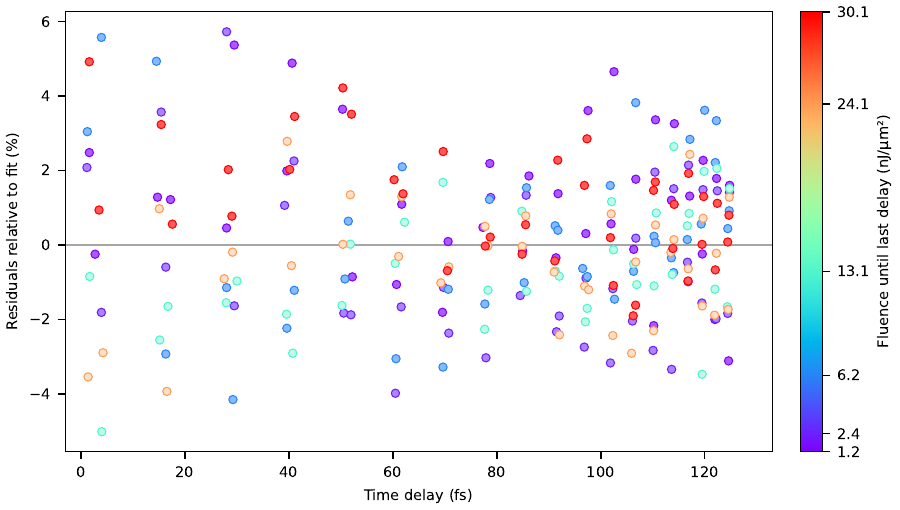}
	\caption{\textbf{Residuals of the fit to the phenomenological model.} 
    The somewhat increased uncertainty at early delays is dominated by low photon statistics since only a small fraction of the \xray{} pulse contributes to diffraction.}
	\label{fig:residuals}
\end{figure}

Seven~free parameters were globally fitted to a total of 228~data points across 6~fluence traces. The fit parameters are listed in Tab.~\ref{tab:fitted_values} together with the residuals of the fit shown in Fig.~\ref{fig:residuals}. The local timing uncertainty $\sigma_t$ was taken into account in the fits by propagating it into the uncertainty of $R$ through the local slope of the signal, yielding an effective uncertainty
\begin{equation}
    \sigma_{\mathrm{eff}}^2 = \sigma_y^2 + \left(\frac{\partial R}{\partial t}\,\sigma_t\right)^2,
\end{equation}
where $\sigma_y$ denotes the statistical uncertainty of the measured intensity ratio $R$. The derivative $\partial R/\partial t$ was evaluated numerically. The resulting $\sigma_{\mathrm{eff}}$ was used to weight the residuals in a least-squares fit. Fig.~\ref{fig:ratio}B shows $1.96\,\sigma_y$ (95\,\% confidence interval) and $\sigma_t$ as error bars. The timing uncertainty becomes noticeable for late time delays where the derivative is already small, thus the propagated timing uncertainty increases the point-wise uncertainty only marginally. Using the measured fluence profile from the (220) reflection or a monochromator scan instead of the linear approximation does not change the fit results, resulting in the same extracted $\tau\approx\ensuremath{19\,\mathrm{fs}}$ timescale.

\newpage


\clearpage

\clearpage

\clearpage

\clearpage

\clearpage

\clearpage

\begin{table}
    \centering
    \caption{\textbf{Fit parameters of the phenomenological model.}
     The model described in the main text and the Materials and Methods section was fitted globally using 7~free parameters to 228~data points across 6~\xray{} fluences. The fit reaches $\chi^2=503.3$ and a reduced $\chi^2_\nu=2.27$. The uncertainties correspond to one-standard-deviation confidence intervals derived from the covariance matrix of the weighted least-squares fit scaled by $\sqrt{\chi^2_\nu}$, using the experimentally determined uncertainties of the Bragg-peak-ratio measurements.\\
    }
    \label{tab:fitted_values}
    \begin{tabular}{l l r}
        \hline
        Parameter & Description & Value \\
        \hline
        $R_0$ & Equilibrium Bragg ratio & \ensuremath{5.96\pm0.13\,\%} \\
        $\Delta R_0$ & Electronic suppression amplitude & \ensuremath{2.13\pm0.12\,\%} \\
        $\tau$ & Timescale of electronic suppression & \ensuremath{18.6\pm2.1\,\mathrm{fs}} \\
        $F_s$ & Phonon restoration saturation fluence & \ensuremath{24.7\pm6.1\,\mathrm{nJ/\mu m^2}} \\
        A & Scattering ratio restoration at full saturation & \ensuremath{1.34\pm0.18\,\%} \\
        $t_0$   &  Time-zero offset & \ensuremath{-1.2\pm2.9\,\mathrm{fs}} \\
        $f$   &  Phonon frequency & \ensuremath{15.1\pm1.5\,\mathrm{THz}} \\
        \hline
    \end{tabular}
\end{table}







\end{document}